\documentclass{iau}
\usepackage{graphicx} 

\title[IAUS291.~~Pulsed $\gamma$-ray from magnetar 1E 2259+586] %% short title %%
{Pulsed $\gamma$-ray emission from magnetar\\ 1E 2259+586} %% full title %%

\author[J. H. K. Wu et al.]  %% short author list %%
{J. H. K. Wu$^1$,
 C. Y. Hui$^2$, R. H. H. Huang$^1$, A. K. H. Kong$^{1}$\thanks{Golden Jade Fellow of Kenda Foundation, Taiwan}, \\
 K. S. Cheng$^3$, J. Takata$^3$, P. H. T. Tam$^1$, E. M. H. Wu$^3$ \\
\and C.-Y. Liu$^1$}

\affiliation{$^1$Institute of Astronomy and Department of Physics, National Tsing Hua University, \\Hsinchu, Taiwan\\ email: {\tt wuhkjason@gmail.com} \\[\affilskip]
$^2$Department of Astronomy and Space Science, Chungnam National University, Daejeon, Republic of
Korea \\ [\affilskip]
$^3$Department of Physics, University of Hong Kong, Pokfulam Road, Hong Kong}

\pubyear{2012}
\volume{291}  %% insert here IAU Symposium No.
\jname{\mbox{Neutron Stars and Pulsars: Challenges and Opportunities after 80 years}}
\editors{J. van Leeuwen, ed.} 
\begin{document}

\maketitle

%% -- Abstract ----------------------------------
\begin{abstract}
Anomalous x-ray pulsars (AXPs) are thought to be magnetars which are young isolated neutron stars with extremely strong magnetic fields of $>10^{14} $Gauss. Their tremendous magnetic fields inferred from the spin parameters provide a huge energy reservoir to power the observed x-ray emission. High-energy emission above 0.3 MeV has never been detected despite intensive search. Here, we present the possible Fermi Large Area Telescope (LAT) detection of $\gamma$-ray pulsations above 200 MeV from the AXP, 1E 2259+586, which puts the current theoretical models of $\gamma$-ray emission mechanisms of magnetars into challenge. We speculate that the high-energy $\gamma$-rays originate from the outer magnetosphere of the magnetar.

\end{abstract}

% add below any authors, subjects and objects for indexing 
%   add more lines if necessary
%   but leave all lines commented out
%\index[author]{LastName1, Initials|textbf}
%\index[author]{LastName2, Initials|textbf}
%\index[subject]{Keyword1}
%\index[subject]{Keyword2}
%\index[object]{Object1}
%\index[object]{Object2}

\firstsection % if your document starts with a section,
\section{Introduction}
Neutron stars are now known to have many different manifestations besides rotation-powered and accretion-powered pulsars. While pulsars typically have a surface magnetic field of $\sim10^{12}$ G, it has been suggested that neutron stars can possess magnetic fields with a strength as high as $\sim10^{15}$ G (Duncan et al. 1992). These highly magnetized neutron stars are called magnetars. The existence of magnetars provides a unique laboratory for exploring the physics of compact objects in the presence of a ultra-strong magnetic field.
Based on current observations and theories, the emissions from magnetars mainly emerge in x-ray energy bands; their broad band spectral shapes can be well described by a blackbody component (with a hard tail probably due to Compton scattering) below 10 keV, which is likely from the magnetarÕs surface, plus a non-thermal component dominating above 10 keV, which is from the magnetosphere (Thompson et al. 2002). On the basis of theoretical models of high-energy emission from magnetars, it is not expected to detect emission above $\sim$ 1 MeV (Thompson et al. 1995).
Although Castro et al. (Castro et al. 2012) have report the $\gamma$-ray emission from CTB 109, which is also well-known for hosting an AXP 1E 2259+586, they concluded that 1E 2259+586 is not likely to be contributing the observed $\gamma$-ray flux. We have done a detail timing analysis based on the timing ephemeris reported in (Icdem et al. 2012), a 5-sigma $\gamma$-ray pulsation from 1E 2259+586 was found. The possible detection of the $\gamma$-ray pulsation suggest that 1E 2259+586 could also contribute part of the $\gamma$-ray flux.

\section{Data analysis}

For the spectral analysis, we used the LAT data between 2008 August 04 and 2011 November 09 (3.5 years of data). To reduce and analyze the data, the Fermi Science Tools v9r23p1 package, available from the Fermi Science Support Center, was used. We used Pass 7 data and selected events in the ÓSourceÓ class (i.e. event class 2) only. In addition, we excluded the events with zenith angles larger than $100^{\circ}$ to greatly reduce the contamination by Earth albedo gamma-rays. The instrumental response functions (IRFs) ÒP7SOURCE V6Ó were adopted throughout the study. Figure 1 shows the binned energy spectrum of CTB 109/1E 2259+586.
As 1E 2259+586 is known to have frequent glitches that are sudden increases in its spin frequency, we only used data taken after the last glitch seen on 2009 February 18 to search for $\gamma$-ray pulsation. By the timing ephemeris report by (Icdem et al. 2012), the period after the microglitch 2 is found to be 6.979060682s, with the aid of the Fermi plug-in for TEMPO2, we assigned a spin phase for each $\gamma$-ray photons with energy greater than 200 MeV and fall within $0.6^{\circ}$ from the AXP position, pulsed $\gamma$-ray emission was found after 120 days since the latest microglitch reported, the folded pulse profile for epoch 2 only is shown in Figure 2 with a 5-sigma pulsation significance using H-test, the 120 days delay of the $\gamma$-ray pulsation could be explained by the decrease of the soft X-ray flux during the above period as the soft X-ray photon would cause photon photon pair creation which eventually prevents those $\gamma$-ray photons to escape from the magnetar. Swift X-ray observation reveal that the soft X-ray flux on 1E 2259+586 during the above period is two times smaller than other epoch. From the folded lightcurve in epoch 2 we can estimate the pulsed fraction to be $\sim$ 30\%. Previous searches on magnetars by Abdo et al. (Abdo et al. 2010) shows no conclusive evidence using 17 months of LAT data, the non-detection on 1E 2259+586 can be explained as Abdo et al. only include LAT photons up to 01 January 2010 ($\sim$55200 MJD), which only covers part of the epoch 2 so that pulsed $\gamma$-ray from 1E 2259+586 is not detectable during the previous search.

\section{Discussion}
If the detection of the pulsed $\gamma$-ray from 1E 2259+586 is genuine, it makes 1E 2259+586 as the first magnetar with GeV $\gamma$-ray emission, it also demonstrates that magnetars are capable to emit GeV $\gamma$-rays similar to canonical $\gamma$-ray pulsars which are powered by the rotational energy; this is indeed predicted more than a decade ago that GeV $\gamma$-rays can originate from the outer magnetosphere of a magnetar (Cheng et al. 2001). However, the acceleration mechanism of the observed pulsed $\gamma$-ray radiations is probably different from the existing model. First, the observed pulsed $\gamma$-ray must be produced far from the magnetarÕs surface to avoid absorption through the one photon pair-creation process, and/or the two photon pair-creation process. Under the circumstance of magnetar, it is likely that the observed pulsed $\gamma$-rays are produced at r $\ge$ 1000 km above the stellar surface. Second, the luminosity of the observed pulsed $\gamma$-ray, L$\gamma$ $\sim 10^{34} erg/s^{-1}$ (assuming isotropic emissions), is well above its spin down power, $L_{sd} \sim 10^{31} erg/s^{-1}$, suggesting that unlike canonical pulsars, the emission process is not powered by its rotational energy loss. Finally, the observational results obtained by x-ray and $\gamma$-ray instruments have suggested that the high-energy emissions of magnetars do not extend beyond several hundred keV (Kuiper et al. 2004; den Hartog et al. 2008; Enoto et al. 2010; Kuiper et al. 2012; Sasmaz Mus et al. 2010, Abdo et al 2010). This suggest that the pulsed $\gamma$-rays should be produced by different mechanism from that of the x-ray emissions.

\begin{figure}[t]
% \vspace*{-2.0 cm}
\begin{center}
 \includegraphics[width=4.2in]{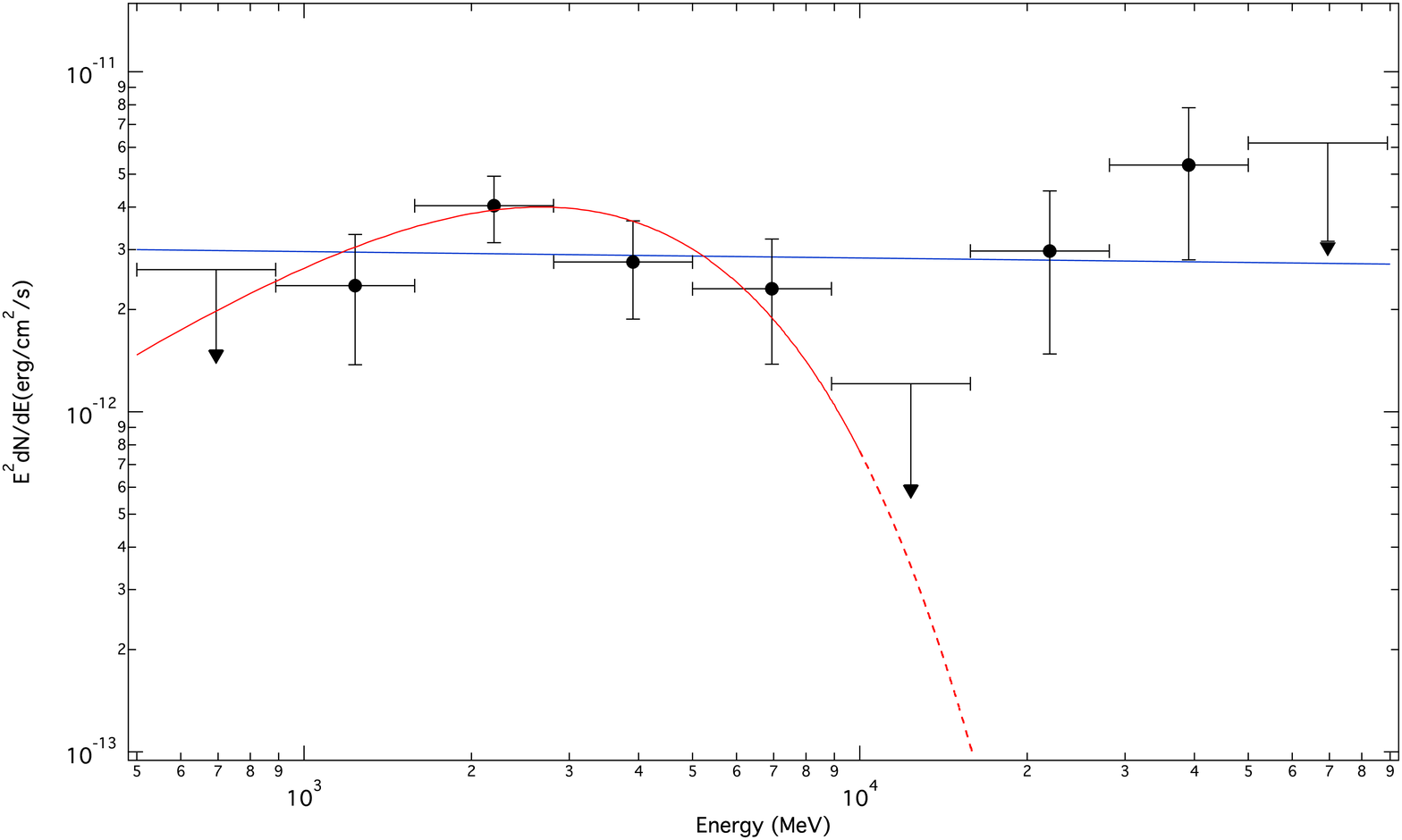} 
% \vspace*{-1.0 cm}
 \caption{Fermi LAT energy spectrum of CTB 109/1E 2259+586. We tried to model the $\gamma$-ray spectrum CTB109/1E 2259+586 by assuming a power law (straight line) or a power law with an exponential cutoff (curved line) model.}
   \label{fig1}
\end{center}
\end{figure}

\begin{figure}[t]
% \vspace*{-2.0 cm}
\begin{center}
 \includegraphics[width=4.2in]{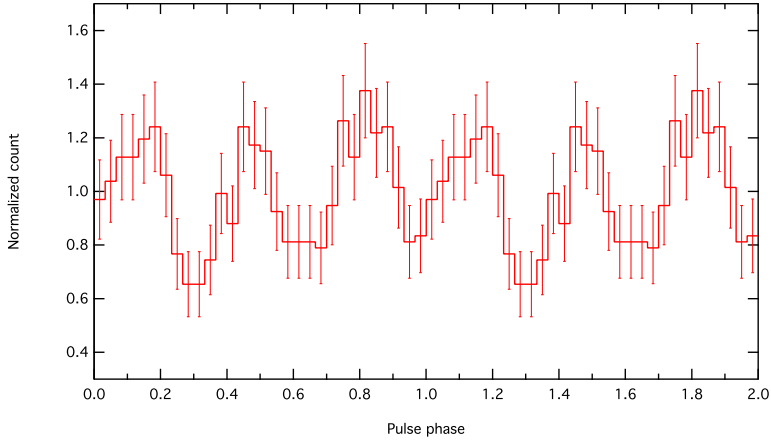} 
% \vspace*{-1.0 cm}
 \caption{Fermi-LAT pulse profile of 1E 2259+586 in epoch 2, $\gamma$-ray photons with energy $>$200 MeV were selected within an aperture radius of $0.6^{\circ}$ around the source position.}
   \label{fig1}
\end{center}
\end{figure}

\end{document}